\documentclass{article}
\usepackage{amsmath}
\usepackage[utf8]{inputenc}
\usepackage{graphicx}
\usepackage[margin=1.2in]{geometry}
\usepackage{cite}
\usepackage{caption}
\title{Hydrodynamic interaction leads to the accumulation of $\textit{Chlamydomonas reinhardtii}$ near a solid-liquid interface}
\author{Chunhe Li, Hongyi Bian, Yateng Qiao, Jin Zhu, Zijie Qu}
\date{2024}
\begin{document}

\maketitle
\section{Abstract}
The physical mechanism of microbial motion near solid-liquid interfaces is crucial for understanding various biological phenomena and developing ecological applications. However, limited works have been conducted on the swimming behavior of $C.\ \textit{reinhardtii}$, a typical ``puller'' type cell, near solid surfaces, particularly with varying and conflicting experimental observations. Here, we investigate the swimming behavior of $C.\ \textit{reinhardtii}$ using a three-dimensional real-time tracking microscopy system both near a solid-liquid interface and in the fluid bulk region. We explore the relationships between the cell density, swimming speed and orientation with respect to the distance from the solid-liquid interface, confirming the phenomenon of $C.\ \textit{reinhardtii}$ accumulation near the solid-liquid interface. Based on the traditional definitions of ``pusher'' and ``puller'' cells, we propose a simplified model consisting of two pairs of mutually perpendicular force dipoles for $C.\ \textit{reinhardtii}$. This model is employed to analyze the complex hydrodynamic interactions between $C.\ \textit{reinhardtii}$ and the solid surface, providing a potential theoretical explanation for the observed accumulation phenomenon at the solid-liquid interface.

\section{Introduction}
Microogranisms achieve locomotion in fluidic environment using various strategies, including rotating rigid helical flagellum \cite{doi:10.1128/jb.182.10.2793-2801.2000,PERSAT2015988}, beating flexible cilia \cite{guasto_oscillatory_2010,annurev:/content/journals/10.1146/annurev-fluid-122414-034606} and constantly deforming their cell bodies \cite{PhysRevFluids.1.053202,Juarez_2010}, to conquer the time-reversibility of the Stokes equation \cite{10.1119/1.10903}. The flow fields generated by these creatures differ dramatically as a result of their swimming modes and are commonly classified as a pusher or puller type \cite{arratia_life_2022}. The difference between these types lies in the direction of flow along the axis of net swimming \cite{drescher_direct_2010,10.1063/1.4718446,doi:10.1073/pnas.1019079108}. Understanding the hydrodynamic interaction between the self-propulsion cells and the ambient fluid is crucial in revealing the physical essence of many biological phenomena, such as the biofilm formation \cite{fei_nonuniform_2020,SANTORE2022102665,doi:10.1126/science.aaq0143}, the spread of intestinal diseases\cite{peng_crb1-associated_2024,scheidweiler_spatial_2024}, and bacteria swarming \cite{10.7554/eLife.64176,PhysRevLett.108.148101}. Moreover, most of these processes occur near a fluid-solid interface. Thus, it is important to study the swimming behavior of microorganisms near solid surfaces.

In previous works, several physical mechanisms have been proposed to elucidate the cell-surface interaction including the wall scattering \cite{kantsler_ciliary_2013,lushi_scattering_2017}, steric interaction \cite{scheidweiler_spatial_2024} and hydrodynamic interaction \cite{yuan_hydrodynamic_2015,PhysRevE.82.056309,Pimponi_Chinappi_Gualtieri_Casciola_2016}. These interactions lead to a surface accumulation of pusher-type cells, like bacterium \emph{E. coli} and sperm \cite{DOMINICK2018588,PhysRevLett.103.078101,ELGETI20101018}. Nevertheless, \emph{Chlamydomonas reinhardtii}, a typical puller-type cell, has also been observed to accumulate near the solid surface. In one of the most recent works, Buchner \emph{et al.} \cite{buchner_hopping_2021} used a 3D tracking microscope to investigate the hopping trajectory of \emph{C. reinhardtii} near the solid surface and conclude that the accumulation is due to long-range interaction. However, the physical interpretation of such interaction remains unclear.

Here, we use a customized 3D tracking scope that actively adjusts the focal plane and microscope stage in real-time to investigate the swimming behavior of \emph{C. reinhardtii} in an unconstrained chamber. Our experimental approach ensures long-term tracking with sufficient statistics for individual behavior analysis. Similar to the existing works, the surface accumulation of the cell and the hopping trajectory are both observed. A double force dipole model is developed to explain the hydrodynamic interaction between the cell and the surface which leads to the accumulation. And the cell distribution in the observation chamber at the steady state is also well explained using a convection-diffusion model.

\section{Result and Discussion}
\subsection{$C.\ \textit{reinhardtii}$ accumulate near solid surface}
\begin{figure}[htbp]
    \centering
    \includegraphics[width = 1\textwidth]{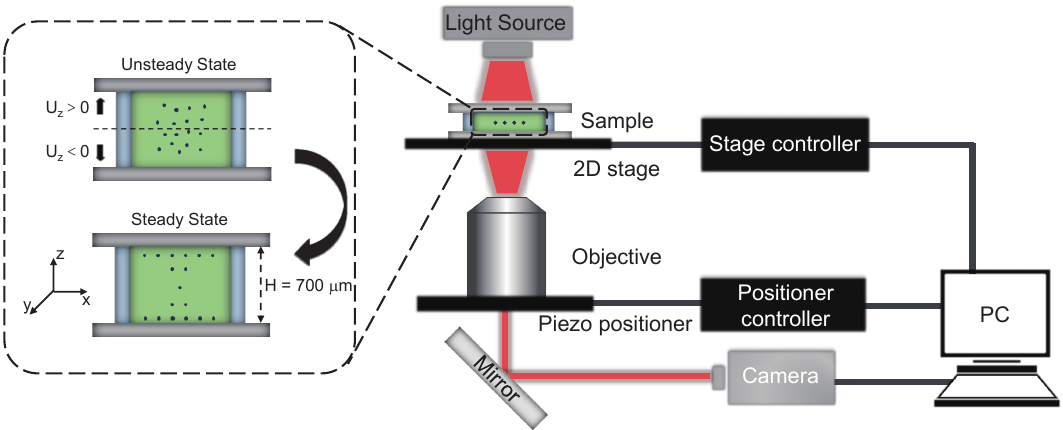}
    \caption{
    Schematic of the experimental apparatus. The cell suspension performing two occasions is placed in a cylindrical chamber and illuminated under red light. Manually adjusting the z-axis height and capturing images to measure the distribution of cells at different distances from the wall. The microscope, duplicated from Qu \emph{et al.}'s study~\cite{qu_changes_2018}, includes a stage controller and a positioner, which work together to record the 3D position of cells.  
    }
    \label{system}
    \end{figure}

To investigate the distribution of $C.\ \textit{reinhardtii}$ density at different distances from the solid-liquid interfaces, the cell suspension is injected in an observation chamber with a height of 700 $\mu m$ (approximately 70 times the body length of the cell $D$ \cite{doi:10.1073/pnas.2206738119}, see Fig. \ref{system}). The other two dimensions of the chamber are approximately 1 cm, so the wall effect from the side is negligible. Cell concentration is measured on two occasions. One is right after the cell suspension is put into the observation chamber. In this case, the cell density is uniform and we define it as the initial state. The other one is when the cell density at different heights of the chamber no longer changes over time, which is the steady state. From our measurement, the typical time for the system to reach the steady state is 30 min. 

\begin{figure}[htbp]
    \centering
    \includegraphics[width = 0.5\textwidth]{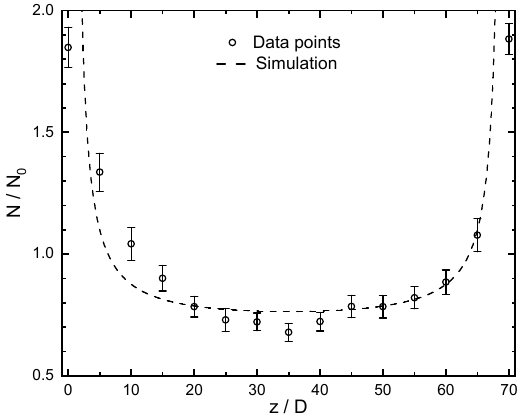}
    \caption{Density distribution of $C.\ \textit{reinhardtii}$ at different distances from the solid-liquid interfaces. Data points indicate the measured distribution which can be expressed as a function of distances to the lower surface ($z / D = 0$). The black dashed line, based on convection-diffusion simulation \cite{berke_hydrodynamic_2008}, fits the data points with $N_0$ = 0.67 and $L$ = 23 $\mu m$. }
    \label{Distribution and Simulation}
    \end{figure}

At the steady state, the number density of the cell is measured and depicted in Fig. \ref{Distribution and Simulation}. (Note: Percoll is added to the suspension to balance the effect of gravity \cite{berke_hydrodynamic_2008}). $z/D$ represents the distance between the cell in the chamber and the lower surface, $z/D = 0$ represents the location near the lower surface, and $z/D = 70$ represents the location near the upper surface. A comparison between suspension with and without Percoll is shown in SI Fig. 1. Obviously, the cell density is much higher near the top and bottom surfaces of the chamber than in the bulk region. Our result is similar to the one from Buchner \emph{et al.} \cite{buchner_hopping_2021}. Notably, although $\emph{E. coli}$ and $C.\ \textit{reinhardtii}$ are known to swim very differently \cite{C7SM02301D,doi:10.1073/pnas.0906586106}, the accumulation phenomenon observed here and previous works \cite{buchner_hopping_2021} shows strong similarity to those studies focusing on $\emph{E. coli}$ cells \cite{berke_hydrodynamic_2008,ahmadzadegan_hydrodynamic_2019}. It has been argued that the reason for $\emph{E. coli}$ being ``trapped'' is the hydrodynamic interaction between a pusher-type force dipole and a solid surface \cite{PhysRevLett.131.158301,doi:10.1073/pnas.2206738119}. However,  $C.\ \textit{reinhardtii}$ has been treated as a typical ``puller-type'' cell, which makes the accumulation phenomenon puzzling. In fact, it has been shown that $C.\ \textit{reinhardtii}$ is repelled from the hard surface either through a direct-contact mechanism \cite{doi:10.1073/pnas.1210548110} or a non-contact interaction \cite{klindt_flagellar_2015}.

\subsection{Swimming behavior near the surface}
\begin{figure}[htbp]
    \centering
    \includegraphics[width = 1\textwidth]{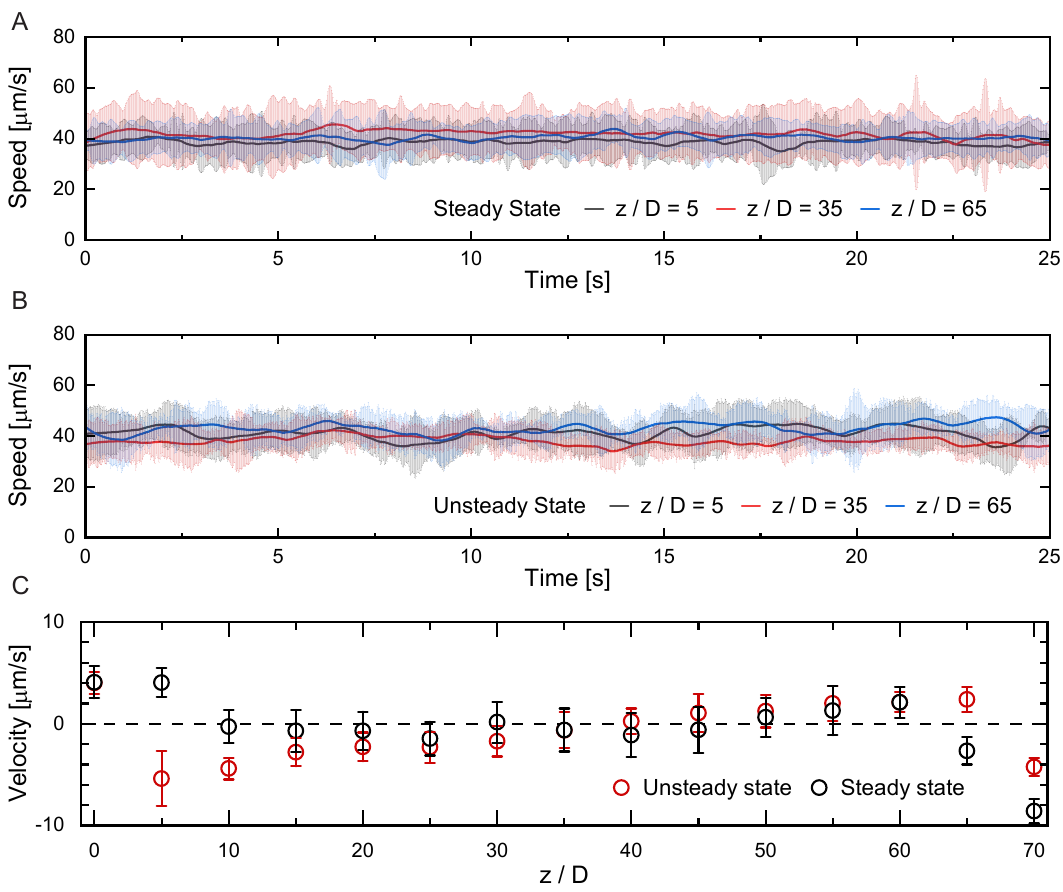}
    \caption{(A and B) Changes in the speed of $C.\ \textit{reinhardtii}$ at different positions in the Chamber over time when the system is in steady state and unsteady state. Ten cells are selected for each position, the solid line is the average speed of cell motion, and the shaded part represents the standard deviation.(C) $z$-velocity component of steady state cells(Black points) and unsteady state cells (Red points) at different distances from the solid-liquid interfaces.}
    \label{Velocity}
    \end{figure}

 To fully understand the swimming behavior of $C.\ \textit{reinhardtii}$, a three-dimensional real-time tracking microscope is developed (see Fig. \ref{system}) to follow individual cells over an extended period of time. Over 30 cells are tracked at different heights and the swimming behavior of each cell is recorded over 25 seconds (see SI Fig. 2). First, we compare the swimming speed at three typical heights for both the steady and unsteady cases. As depicted in Fig. \ref{Velocity}A and B, the average swimming speed is almost the same. This indicates that neither the cell density nor the existence of the solid boundary influences the absolute swimming speed of the cell. Next, the vertical component of the velocity, $U_z$ (velocity along the $z$ axis) is averaged and shown in Fig. \ref{Velocity}C. Here, an interesting trend is found. In the unsteady state (right after the suspension is injected), $U_z$ increases linearly with height and is close to zero in the middle of the chamber. This implies that the cell quickly migrates towards the solid boundary right after the suspension is injected. Such “attraction” scales linearly with distance. It is also worth noting that $U_z$ is positive at the bottom surface and negative at the top, violating the overall trend. However, this can be well explained by the fact that the cell is bounced back from the surface as mentioned by different studies \cite{ostapenko_curvature-guided_2018, doi:10.1073/pnas.1210548110}. After the system evolves to the steady state, $U_z$ is close to zero at different heights (Fig. \ref{Velocity}C, black markers) except at the surfaces where the bouncing mechanism still exists.

\subsection{Composite force dipoles hypothesis}
Where does such ``attraction'' come from? Here, we propose a hypothesis that the hydrodynamic interaction between \textit{$C.\ \textit{reinhardtii}$} and the fluid is a combination of a simple “puller-type” force dipole and a “pusher-type” force dipole. As it has been previously studied, \textit{$C.\ \textit{reinhardtii}$} swims by periodically beating a pair of cilia \cite{polin_chlamydomonas_2009,PhysRevE.107.014404}. The model that the cell is a “puller” roots in the fact that over one breaststroke cycle, the cell is being pulled by the cilia in front of it. However, during the initial part of this process, both cilia are swung to the side of the cell, exerting little pulling force on the cell body. This still introduces a strong interaction with the fluid and we believe it can be treated as a “pusher-type” force dipole. The later part of the breaststroke cycle still introduces a “puller-type” interaction as it has been investigated for a long time \cite{guasto_oscillatory_2010,klindt_flagellar_2015}.

\begin{figure}[htbp]
    \centering
    \includegraphics[width = 1\textwidth]{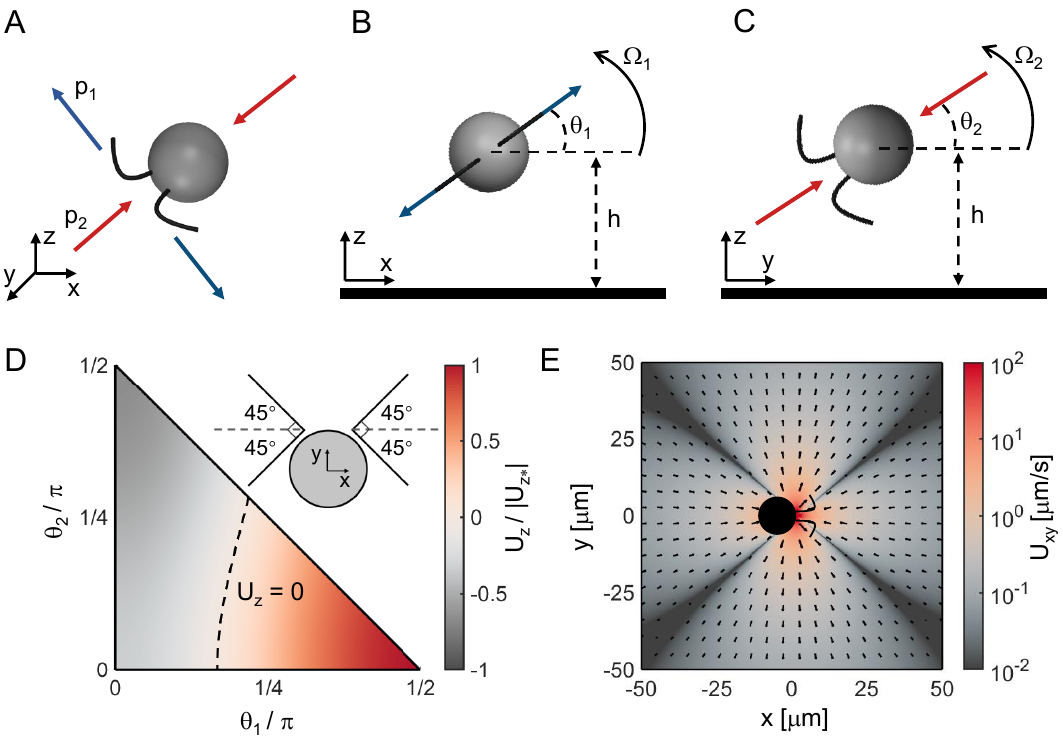}
    \caption{(A) Schematic diagram illustrating the assumption of composite force dipoles. Force dipoles $p_1$ and $p_2$ are in a perpendicular configuration. $p_1$ behaves akin to a “pusher,” while $p_2$ behaves akin to a “puller.” (B) The angle of $p_1$ and (C) $p_2$ with respect to the plane and the angular velocity of rotation, when the distance of the cell from the wall is $h$. (D)  Simplified depiction of $C.\ \textit{reinhardtii}$ swimming, with flagellar beating within a 90° range. Under the assumption of a length-to-diameter ratio of 2, variations in fluid velocity around the cell for different angles between the cell and the plane are illustrated. Color is normalized by the local velocity and the maximum magnitude of velocity. (E) xy-plane simulation velocity field near $C.\ \textit{reinhardtii}$ under the “pusher” and “puller” types. Arrow lengths and the color represent the logarithmic scaling of local velocity and maximum magnitude of velocity normalized.}
    \label{Hypothesis}
    \end{figure}

We model \textit{$C.\ \textit{reinhardtii}$} as a pair of perpendicularly located “pusher” dipole, $p_1$ and a “puller”, $p_2$ (see Fig. \ref{Hypothesis}A). Because we study the swimming behavior of the cell for several tenths of seconds, many cycles are achieved. This time-averaged model is validated. Mathematically, the flow field induced by a Stokeslet dipole near a solid surface can be expressed as \cite{berke_hydrodynamic_2008}:  
\begin{equation}
U_z (\theta_1,\theta_2,h)=-\frac3{64\pi\eta h^2}[p_1(1-3\mathrm{sin}^2\theta_1)-p_2(1-3\mathrm{sin}^2\theta_2)]
\label{Uz}
\end{equation}
where $p_i$ is the strength of the force dipole, $\theta_i$ is the angle between the force dipole direction and the parallel plane of the wall, $h$ is the distance between the force dipole and the surface, and $\eta$ is the fluid viscosity. Moreover, the strength of force dipole $|p|$ scales as $\eta{U}{L}^2$ \cite{Lauga_2009}, where U is the swimming speed of the cell, and L is the characteristic length scale. In our future analysis, we choose $L$ = 10 $\mu m$, $U$ = $60\,\mu m /s$ and $\eta$ = $2.98\times10^{-3} Pa\cdot s$. The geometry of the cell is from the previous studies \cite{10.7554/eLife.64176}, the swimming speed comes from our measurement and $\eta$ is the viscosity of the buffer (see SI Fig. 3) at room temperature.

It has been shown from previous studies that the beating velocity of the cilia is much higher during the middle of a complete stroke \cite{Wei_Dehnavi_Aubin-Tam_Tam_2021} when most of the hydrodynamic interaction occurs. The cilia sweep through an area covered from $-45^\circ$ to $45^\circ$ (shown in Fig. \ref{Hypothesis}D) to generate most of the propulsion \cite{friedrich_flagellar_2012}. With these results, we assume that the characteristic length scale for the “pusher” is twice compared with the “puller” while the characteristic velocity for both dipoles is the same. Assuming the length of single cilia and cell diameter are both 10 $\mu m$, we deduce that the ratio of characteristic length scale between “pusher” and “puller” is $(D_{cell}+2\times L_{cilia} \times\cos45^{\circ})/(D_{cell}+L_{cilia}\times\cos45^{\circ} ) \approx 1.414 $(diameter of cell and length of cilia are both 10 $\mu m$). Hence, the strength of $p_1$ is approximately twice greater than that of $p_2$. 

With this assumption, we first calculate the flow field as a result of the two force dipoles using the equation 
\begin{equation}
\boldsymbol U(\boldsymbol r)=\frac {p_1-p_2}{8\pi\eta r^3}\left[1-3\sin^2\theta\right]\boldsymbol r
\label{U}
\end{equation} (shown in Fig. \ref{Hypothesis}E). The result is similar to the ones measured experimentally \cite{drescher_direct_2010} and simulated \cite{lushi_scattering_2017}. Since we argue that this is a time-averaged result, some features such as the vortex of the flow field are missing when compared with the previous works. Next, we use Equ. \ref{Uz} to investigate the velocity along the $z$ axis as a function of $\theta_i$. As depicted in Fig. \ref{Hypothesis}D, when $\theta_1$ is in the range of $0$ to $\pi/6$, regardless of the value of $\theta_2$, $U_z$ is always negative; when $\theta_1$ is greater than $\pi/4$, $U_z$ is always positive. Also, it is noted that since $p_1$ and $p_2$ are always perpendicular to each other,  $0 \leq \theta_1 + \theta_2 \leq \pi / 2$ should always be satisfied which leads to a higher probability of $U_z$ being negative. This indicates that cells near the wall are more likely to be “attracted” as a result of fluid flow, contributing to the occurrence of wall accumulation phenomena. Note: whether the cell is being attracted or repelled is independent of height because $h$ only contributes to the strength of the interaction (Equ. \ref{Uz}), not the sign.

Besides, when a single cell approaches the wall, disturbances in the surrounding flow field can result in the cell being influenced by non-zero velocity gradients, leading to a cell body rotation near the wall \cite{Lauga_2009}. This effect can be expressed by the following equation:
\begin{equation}
\Omega_z(\theta,h)=-\frac{3p\cos\theta\sin\theta}{64\pi\eta h^3}\left[1+\frac{(\gamma^2-1)}{2(\gamma^2+1)}(1+\sin^2\theta)\right]
\end{equation}
where $\gamma$ is the aspect ratio of the dipole. Due to the linearity of the low $Reynolds$ number Stokes flow \cite{berke_hydrodynamic_2008}, we treat this rotational influence on $\theta_1$ and $\theta_2$ separately. The cilia sweep through an area covered from $-45^\circ$ to $45^\circ$ (shown in Fig. \ref{Hypothesis}D) to generate most of the propulsion. Thus, $\gamma$ for the “pusher” and “puller” are taken as $2$ and $1/2$ according to the simplified geometric model. Also, $\gamma$ for the “pusher” and “puller” are taken as $2$ and $1/2$ because of the geometry of the system. Through calculations, we find that for $0 < \theta_1 < \pi/2$, $\Omega_1$ is always negative, leading $\theta_1$ to decrease, and for $0 < \theta_2 < \pi/2$, $\Omega_2$ is positive, leading $\theta_2$ to increase. Again, $h$ contributes only to the strength of the interaction rather than the direction.

\begin{figure}[htbp]
    \centering
    \includegraphics[width = 1.0\textwidth]{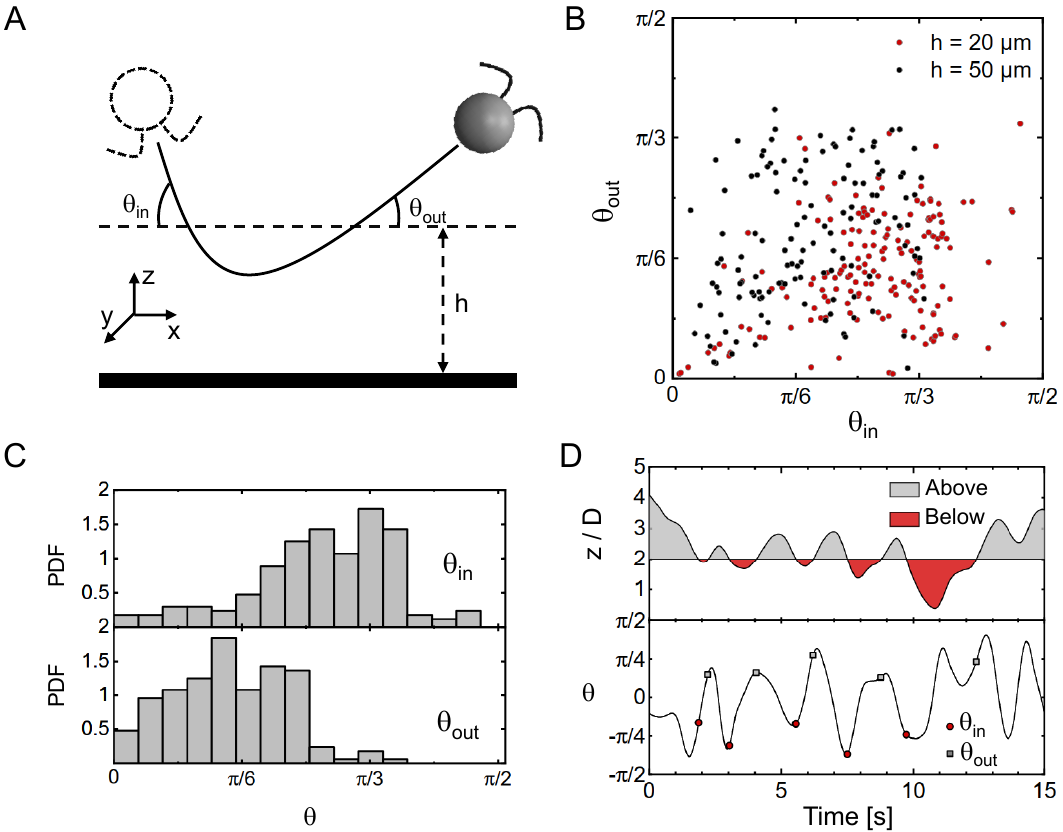}
    \caption{(A) Schematic of the approaching and leaving angles of $C.\ \textit{reinhardtii}$ during swimming near the wall surface . (B) Distribution of $\theta_{in}$ and $\theta_{out}$ for $C.\ \textit{reinhardtii}$ away from ($h$ = 50 $\mu m$) and  near ($h$ = 20 $\mu m$) the surface. (C) The histogram indicates that in the case of $h$ = 20 $\mu m$, approaching angles $\theta_{in}$ are biased around $\pi / 3$, while receding angles $\theta_{out}$ are biased around $\pi / 6$. (D) One typical cell swimming trajectory which has approached and left the surface ($h$ = 20 $\mu m$) many times. The absolute value of $\theta_{in}$ is more likely larger than $\theta_{out}$.} 
    
    \label{Angles}
    \end{figure}

Due to the limitation of our visualization tool, $\theta_1$ cannot be measured directly from the experiment, however, $\theta_2$ can be resolved from the swimming trajectory. It is worth noting that $\theta_2$ is related to $p_2$, so the corresponding experimental results can be used to validate our hypothesis of a new force dipole. To best see this result, we investigate the cell swimming trajectories at different heights, $h$ = 20 $\mu m$ and $h$ = 50 $\mu m$. As depicted in Fig .\ref{Angles}A, we define the smaller angle between the swimming trajectory and the $x$O$y$ plane when the cell is approaching the surface as $\theta_{in}$ and the same angle when the cell is leaving the surface as $\theta_{out}$. We measure the trajectories of 75 cells near this region and plot the distribution of $\theta_{in}$ and $\theta_{out}$ in Fig. \ref{Angles}B and C). For $h$ = 20 $\mu m$,  $\theta_{in}$  tends to be biased around the $\pi / 3$ and $\theta_{out}$ around the $\pi / 6$, however, for $h$ = 50 $\mu m$, the distribution is less biased. With our 3D real-time tracking technique, the swimming trajectory of a typical cell (starting at $h$ = 20 $\mu m$) is shown in Fig .\ref{Angles}D. The result aligns with past research that when $C.\ \textit{reinhardtii}$ swims toward a wall, it will tend to reorient perpendicular to the wall\cite{Lauga_2009}. We analyze a subset of cells within the experimental sample that are in contact with the solid surface. When contacting with the solid surface, $C.\ \textit{reinhardtii}$ scatters at a smaller angle, leading to the differences between the distributions of  $\theta_{in}$ and $\theta_{out}$.

\subsection{Convection-diffusion model for cell distribution}
The double force dipole model well explains the mechanism for $C.\ \textit{reinhardtii}$ being attracted by solid surface. With this result, we further investigate the cell distribution at the steady state. As it is previously discussed, the cell distribution in the observation chamber is uniform right after the suspension is injected and evolves to the steady state when more cells are located near the surfaces (Fig. \ref{Distribution and Simulation}). We believe such an equilibrium state is the result of balanced convection and diffusion, which is illustrated with the following equation:

\begin{equation}
  \frac{\partial}{\partial h}(NU_z)=D\frac{\partial^2N}{\partial h^2}.
   \label{3}
\end{equation}
Here, $D$ is the diffusion coefficient of the cell, As mentioned earlier, we know that $\theta_1$ and $\theta_2$ tend to become 0 and $\pi / 2$ near the wall. Thus, we further simplify the velocity expression in Equ. \ref{4} to: 
\begin{equation}
 U(h)=-\frac3{64\pi\eta h^2}(p_1-2p_2)=-\frac{1.5p_1}{64\pi\eta h^2}   
 \label{4}
\end{equation}
    Finally, by solving Equ. \ref{3}, we obtain the expression for the distribution of cell density with respect to height: 
\begin{equation}
   \frac{N(h)}{N_0}=\exp\left[L\left(\frac1h+\frac1{H-h}\right)\right]
\end{equation}
 where $N_0$ is related to the number of cells in the system, and $L_{}\approx23\mu m$ represents the length scale of the hydrodynamic interaction. 

We estimate the force dipole moment generated by $C.\ \textit{reinhardtii}$ during swimming using the following equation: $p=64\pi\eta DL/3$, under conditions of water at 20°C, the diffusion coefficient of cells is approximated to be the same as that of particles with a diameter of 10 $\mu m$. The estimated result obtained was $p \approx 12\,\mathrm{pN}\,\mu\mathrm{m}$, which is close to and slightly smaller than the value estimated through previous theoretical approaches $p \approx 100\,\mathrm{pN}\,\mu\mathrm{m}$. This is possibly due to the periodic swimming behavior of $C.\ \textit{reinhardtii}$, and the force dipole moment estimated from the experimental data is the average over one period, which is naturally smaller than the maximum value.

\section{Conclusion and discussion}
In this work, we investigate the accumulation phenomenon of $C.\ \textit{reinhardtii}$ near solid-liquid interface. By tracking individual cells, we find a similar ``hopping'' trajectory of $C.\ \textit{reinhardtii}$ near surfaces as previously reported by Buchner \emph{et al.} \cite{buchner_hopping_2021}. We propose a double-perpendicular-force dipole model which is different from Klindt \emph{et al.}'s hypothesis \cite{klindt_flagellar_2015} to explain the hydrodynamic interaction between the boundary and the cell. $C.\ \textit{reinhardtii}$, the well-known puller cell, may exert a stronger pusher-type interaction with the ambient fluid, leading to surface accumulation. A convection-diffusion model is developed to explain the distribution of cells in the observation chamber at the steady state. Nonetheless, a closer examination of the cilia motion is missing in our work and the rotational motion along the axis of swimming cannot be measured due to the limitation of our experimental system. Additionally, the dipole moment used in the hypothesis is a rough estimation, more detailed experimental and theoretical studies are needed. In all, we propose a new perspective on understanding the swimming behavior of $C.\ \textit{reinhardtii}$ near the solid surface and a model for the cell accumulation phenomenon.

\section{Acknowledgement}
We thank Dr. Li Zeng for kindly gifting us the cells. We thank Hepeng Zhang and Shuo Guo for the comments and discussions on the paper content. Our work is funded by This work is funded by NSFC 12202275, STCSM 22YF1419800.

\section{Methods}
\subsection{Cell Culture}
$C.\ \textit{reinhardtii}$ cells (GY-D55) used in the experiments were obtained from Shanghai Guangyu Biological Technology Co., Ltd. The species were cultured in TAP (Tris-Acetate-Phosphate) Broth during the exponential growth period for 10 d to reach the cell density of 1e6 cells/mL counted by a hemocytometer (Watson 177-112C) and then exposed to the 12h-12h light/dark cycle. The temperature was 25 ℃ no matter the light cycle or the dark cycle with the illumination between 2000 and 6000 lux using white LEDs. 
\subsection{Chamber preparation}
Circular chambers made of Polydimethylsiloxane (PDMS) were manufactured by mixing and thoroughly stirring the base and curing agent in a 10:1 mass ratio. After mixing, the mixture was placed in a vacuum chamber for degassing and then carefully pouring between two slides separated by seven spacers, each of which was 100 $\mu m$ thick. All parts were heated in an oven at 80 °C for 2 hours to complete the curing of PDMS. Removing the PDMS film from the oven and cutting neatly for subsequent experiment. At last, the PDMS film was perforated using a 1mm diameter punch.
\subsection{Tracking Microscopy}
A mixture of 100 $\mu l$ of $C.\ \textit{reinhardtii}$ and Percoll (volume ratio of about 1:1) is added to the chamber with a glass substrate with a pipettor, and the chamber is then sealed with a cover slide to ensure that no air is present inside.

3D real-time tracking technology was used to record the three-dimensional swimming trajectory of individual $C.\ \textit{reinhardtii}$ for a long time (about 30 seconds). The cells were viewed by using an inverted microscope model Nikon ECLIPSE Ti2-U with PX.EDGE 5.5-sCMOS camera.A two-dimensional displacement stage was used to track the movement of cells in the direction of parallel glass plates. The piezoelectric displacement stage (Pi) was used to record the movement of the cells in the vertical direction and controls the position of the focal plane to ensure that the cells were always in view. Through the algorithm written by C++ language, the images were obtained at 60fps, 300X300 pixels. And the data of the two-dimensional and the piezoelectric displacement stage were recorded at the same time. The real-time three-dimensional motion trajectory of the cell was finally obtained in the specified time. 
\clearpage

\clearpage
\newpage
%\bibliographystyle{unsrt}
%\bibliography{ref}

\end{document}

\maketitle

\begin{figure}[htbp]
    \centering
    \includegraphics[width=1\textwidth]{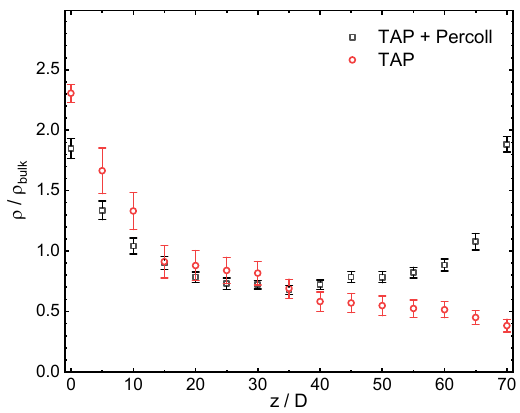}
    \caption{The cell density distribution was compared before and after adding percoll into TAP medium at different distances from the wall.}
    \label{fig:cell_density}
\end{figure}

\begin{figure}[htbp]
    \centering
    \includegraphics[width=1\textwidth]{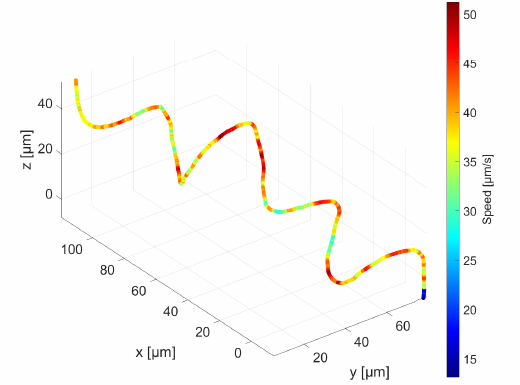}
    \caption{Three-dimensional trajectory of $C. \textit{reinhardtii}$ using the 3D real-time tracking microscope, and colorbar represents the cell swimming speed.}
    \label{fig:3D_trajectory}
\end{figure}

\begin{figure}[htbp]
    \centering
    \includegraphics[width=1\textwidth]{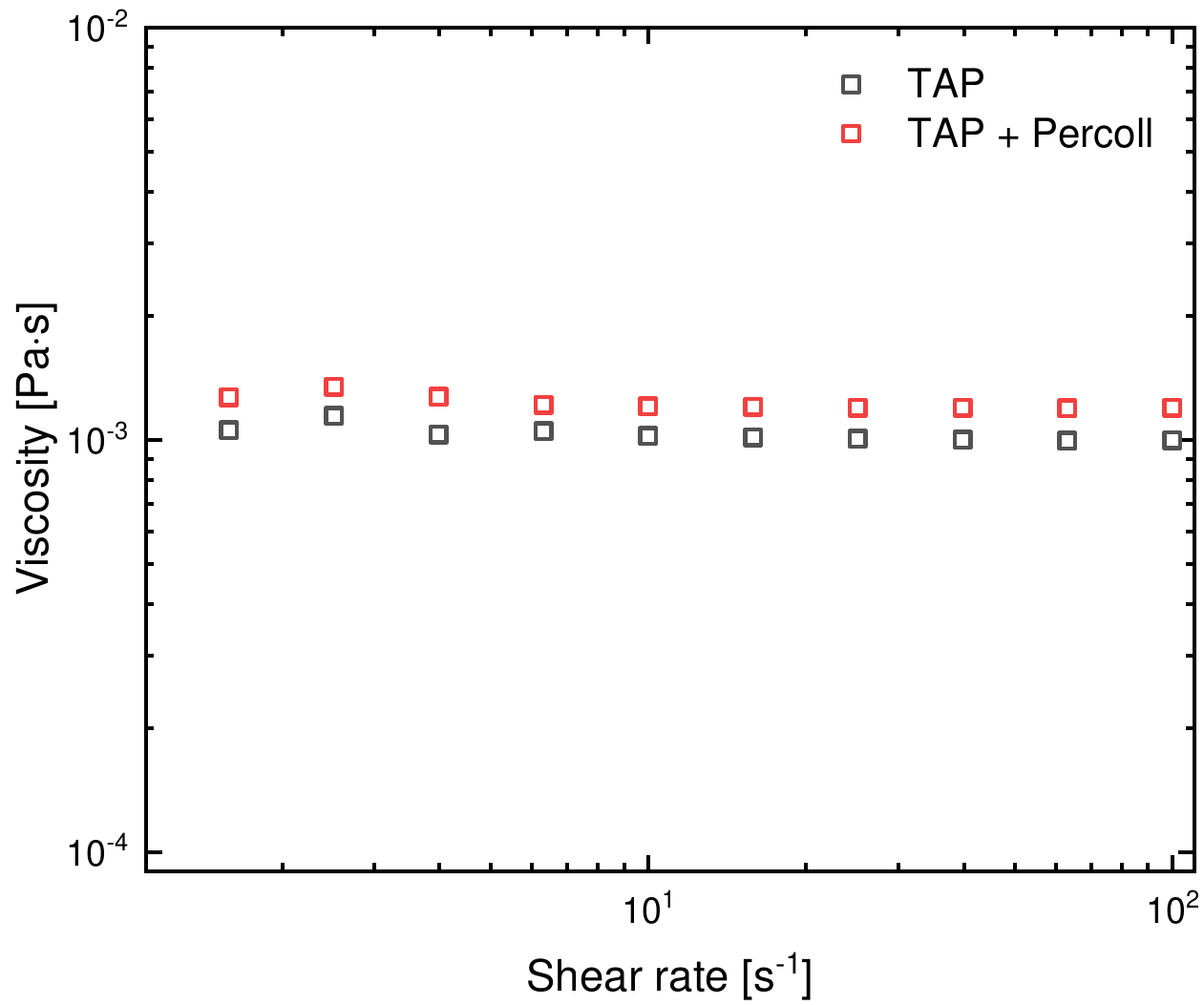}
    \caption{Comparison of solution viscosity before and after adding percoll into TAP medium at different shear rates.}
    \label{fig:viscosity}
\end{figure}

\begin{figure}[htbp]
    \centering
    \includegraphics[width=1\textwidth]{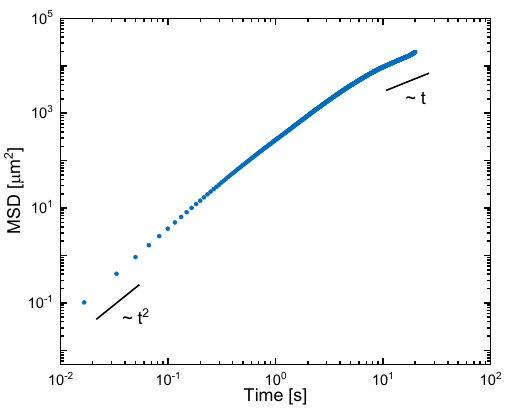}
    \caption{Mean-squared displacement (MSD) of cells. The MSD exhibits a ballistic regime ($\mathrm{MSD}\propto t^2$) at early times and approaches a diffusive regime ($\mathrm{MSD}\propto t$) at later times.}
    \label{fig:MSD}
\end{figure}